\documentclass[aps,pra,reprint,superscriptaddress,showpacs,eqsecnum]{revtex4-1}

\usepackage{graphicx,color}
\usepackage{amsmath}
\usepackage{amssymb}
\usepackage{bm}

\renewcommand{\Re}{\mathop{\text{Re}}\nolimits}

\newcommand{\Tr}{\mathop{\text{Tr}}\nolimits}
\newcommand{\ket}[1]{|{#1}\rangle}
\newcommand{\bra}[1]{\langle{#1}|}

\newcommand{\bracket}[2]{\langle#1|#2\rangle}
\newcommand{\Det}{\mathop{\text{Det}}\nolimits}

\definecolor{dgreen}{rgb}{0,0.5,0}

\definecolor{delete}{cmyk}{0.5,0,0,0}
\definecolor{deletem}{cmyk}{0,0.5,0,0}
\definecolor{deletey}{cmyk}{0.1,0.1,1,0}

\begin{document}


\title{Relevance of Bose-Einstein Condensation to the Interference of\\ Two Independent Bose Gases}



\author{Mauro Iazzi}
\email{mauro.iazzi@sissa.it}
\affiliation{International School for Advanced Studies (SISSA), via Beirut 2-4, I-34014 Trieste, Italy}
\author{Kazuya Yuasa}
\email{yuasa@aoni.waseda.jp}
\affiliation{Waseda Institute for Advanced Study, Waseda University, Tokyo 169-8050, Japan}


\date[]{March 14, 2011}

\begin{abstract}
Interference of two independently prepared ideal Bose gases is discussed, on the basis of the idea of \textit{measurement-induced interference}. It is known that, even if the number of atoms in each gas is individually fixed finite and the symmetry of the system is not broken, an interference pattern is observed on each single snapshot.
The key role is played by the Hanbury Brown and Twiss effect, which leads to an oscillating pattern of the cloud of identical atoms.
Then, how essential is the Bose-Einstein condensation to the interference?
In this work, we describe two ideal Bose gases trapped in two separate 3D harmonic traps at a finite temperature $T$, using the \textit{canonical ensembles} (with fixed numbers of atoms). 
We compute the full statistics of the snapshot profiles of the expanding and overlapping gases released from the traps.
We obtain a simple formula valid for finite $T$, which shows that the average fringe spectrum (average fringe contrast) is given by the \textit{purity} of each gas.
The purity is known to be a good measure of condensation, and the formula clarifies the relevance of the condensation to the interference.
The results for $T=0$ previously known in the literature can be recovered from our analysis.
The fluctuation of the interference spectrum is also studied, and it is shown that the fluctuation is vanishingly small only below the critical temperature $T_c$, meaning that interference pattern is certainly observed on every snapshot below $T_c$.
The fact that the number of atoms is fixed in the canonical ensemble is crucial to this vanishing fluctuation.
\end{abstract}


\maketitle

\section{Introduction}
One of the spectacular phenomena in the physics of ultracold atomic gases is interference.
When two independently prepared Bose-Einstein condensates (BECs) are released and overlap, interference fringes are observed between them \cite{ref:InterferenceBEC}.
This is by no means a trivial phenomenon.
Imagine two sources independently emitting particles toward a screen.
Accumulation of the particles on the screen does not normally result in an interference pattern, since the relative phase between the two wave functions originating from the two independent sources is not well defined in general, which is crucial in Young-type interference experiments.

The simplest description of the interference of independent BECs is based on the \textit{spontaneous symmetry breaking} \cite{ref:BEC-Stringari}: the $\text{U}(1)$ symmetry of the system is spontaneously broken upon condensation and the two gases individually acquire definite phases.
As a result, the relative phase between the gases becomes well defined, which enables them to exhibit interference.
The symmetry breaking, however, would be valid only approximately, since the actual gases in typical interference experiments consist of finite numbers of atoms.
In particular, if the number $N$ of atoms in each gas is precisely fixed and its phase is completely uncertain, interference would not be expected between two such gases in the way to understand the Young-type interference.

Javanainen and Yoo, however, showed in their seminal paper \cite{ref:JavanainenYoo} that, even with such gases with fixed numbers of atoms, an interference pattern can be observed on \textit{each snapshot photo} of the overlapping gases.
Notice that many atoms in the cloud are recorded on a photo at once.
The indistinguishability of the identical bosonic atoms induces quantum correlations among them, which result in a nonuniform distribution of the atoms, in particular, a sinusoidal interference pattern.
The density profiles differ from snapshot to snapshot and the appearance of an interference pattern is not definitely certain.
According to the numerical simulation by Javanainen and Yoo in \cite{ref:JavanainenYoo}, however, sinusoidal patterns are very \textit{typical} among all possible snapshot profiles and interference is \textit{almost certainly} observed on every snapshot (see also \cite{ref:PolkovnikovEPL}).
The interference patterns shift randomly from snapshot to snapshot and the superposition of many snapshots results in an image with no interference.
This is due to the \textit{independence} of the two gases with no phase correlation.
One of the interference patterns and one definite relative phase are selected by taking a photo, i.e., by \textit{measurement}, and such interference revealed on a snapshot is called \textit{measurement-induced interference} \cite{ref:MI,ref:MI-int,ref:MI-Rev}.

Interference of BECs has been attracting renewed attentions these years, and a variety of interference experiments have been reported \cite{ref:IntThermal,ref:Hadzibabic,ref:Schmidmayer}.
In particular, the fluctuations of the interference patterns are shown to provide fruitful information to probe complex many-body states of ultracold trapped atoms \cite{ref:PolkovnikovEPL,ref:Hadzibabic,ref:Schmidmayer,ref:ProbeCorrelationAltman,ref:Boston,ref:PatrickRathZwerger}, and the statistics of interference patterns has become an interesting subject to study.

The main purpose of the present work is to clarify the relevance of the Bose-Einstein condensation to the interference of two independent BECs, on the basis of the idea of measurement-induced interference.
As shown by Javanainen and Yoo, higher-order correlations play a crucial role for the appearance of the interference fringes.
Then, how about the condensation?
\textit{How is Bose-Einstein condensation essential to the interference?}
This is the question we wish to address in the present paper.
Recently, Rath and Zwerger have shown by a simple argument that the visibility of the interference is directly related to the condensation fraction \cite{ref:PatrickRathZwerger}.
In this paper, we provide another evidence of this relationship between the interference and the condensation.

We consider the following setup.
We prepare two independent ideal gases of bosonic atoms trapped in two spatially separated 3D harmonic traps at a finite temperature $T$.
Each gas contains exactly $N$ atoms separately and is described by a \textit{canonical ensemble} with the fixed number of atoms.
These gases are then released from the traps, expand in 3D free space, and overlap.
We are interested in the interference patterns appearing on snapshot photos of the cloud of overlapping gases.

To carry out our analysis, we compute the characteristic functional of the statistics of the snapshot profiles of the cloud, valid for the whole range of temperature $T$ (across the critical temperature $T_c$) for a large number of atoms $N$.
In particular, we find that the average strength of the interference spectrum (Fourier spectrum of the density profile) over all snapshots, which is related to the average visibility of the interference pattern, is simply given by the ``purity'' of each gas: the larger is the purity of the gas, the higher is the contrast of the interference, and no interference is expected above the critical temperature $T_c$.
The purity is known to be a good measure of condensation \cite{ref:CondPurity}, because it is large when only a few states are macroscopically occupied and approaches $1$ when only one state is populated. This shows that Bose-Einstein condensation is relevant to the interference.

Furthermore, we see that the fluctuation of the interference spectrum is vanishingly small at any temperature below the critical temperature $T<T_c$, while the fluctuation abruptly changes at the critical temperature $T_c$ and becomes nonvanishing above $T_c$.
The interference pattern with fringe contrast depending on the purity of the gases is \textit{typical} among all possible profiles and is \textit{certainly observed on every snapshot} below the critical temperature $T<T_c$.
It is shown that the canonical ensemble, in which the number of atoms is fixed, is crucial to the vanishing fluctuation.

This paper is organized as follows.
We set up tools to study the statistics of the snapshot profiles in Sec.\ \ref{sec:Tools}, which are shown to be essentially the same as the ones employed in \cite{ref:PolkovnikovEPL,ref:Boston,ref:PatrickRathZwerger}.
In Sec.\ \ref{sec:SingleGas}, the characteristic functional characterizing the canonical ensemble of a single gas with a fixed number of noninteracting atoms in a harmonic trap is given, which is the key ingredient in the present analysis.
From this, in Sec.\ \ref{sec:Z}, we derive the characteristic functional of a pair of such harmonic clouds and compute the full statistics of the snapshot profiles after the two gases are released and overlap.
The average and the covariance of the fringe spectrum are then analyzed in detail in Sec.\ \ref{sec:InterferenceFluctuation}, obtaining the concise formula for the average spectrum given by the purity, and the fluctuation of the fringe spectrum is investigated as a function of the temperature.
Finally, a summary of the work is given in Sec.\ \ref{sec:Conclusion}, and some details of the calculations, concerning the derivation of the characteristic functional of the snapshot profiles, the treatment of the canonical ensemble, and the estimation of the purity and the other relevant quantity,\ are presented in Appendices \ref{app:GeneFunc}--\ref{app:Purity}.

\section{Statistics of Snapshot Profiles}
\label{sec:Tools}
First of all, we setup some mathematical tools, which are used in the following analysis.
Suppose that there are a large number of identical bosonic atoms and one takes a photo of the cloud: the positions of the $N$ atoms are recorded at once on the snapshot.
The probability of finding the $N$ atoms at positions $\{\bm{r}_1,\ldots,\bm{r}_N\}$ at an instant $t$ is given by
\begin{align}
&P_t^{(N)}(\bm{r}_1,\ldots,\bm{r}_N)
\nonumber\\
&\qquad
=\frac{1}{N!}\langle
\hat{\psi}^\dag(\bm{r}_1)\cdots\hat{\psi}^\dag(\bm{r}_N)
\hat{\psi}(\bm{r}_N)\cdots\hat{\psi}(\bm{r}_1)
\rangle_t,
\end{align}
where $\hat{\psi}(\bm{r})$ is the field operator of the bosonic atom, satisfying the canonical commutation relations
\begin{equation}
[\hat{\psi}(\bm{r}),\hat{\psi}^\dag(\bm{r}')]=\delta^3(\bm{r}-\bm{r}'),\quad
\text{etc.},
\end{equation}
and $\langle\cdots\rangle_t$ denotes the expectation value estimated in the state of the cloud at time $t$.
This probability is normalized to unity as 
\begin{equation}
\int
\prod_{\ell=1}^N d^3\bm{r}_\ell\,
P_t^{(N)}(\bm{r}_1,\ldots,\bm{r}_N)=1.
\end{equation}
The probability to find $M$ atoms among $N$ at $\{\bm{r}_1,\ldots,\bm{r}_M\}$ is given by 
\begin{align}
&P_t^{(M)}(\bm{r}_1,\ldots,\bm{r}_M)
\nonumber\\
&\qquad
=\frac{(N-M)!}{N!}\langle
\hat{\psi}^\dag(\bm{r}_1)\cdots\hat{\psi}^\dag(\bm{r}_M)
\hat{\psi}(\bm{r}_M)\cdots\hat{\psi}(\bm{r}_1)
\rangle_t,
\end{align}
where the normalization is such that 
\begin{align}
&\int d^3\bm{r}_\ell\,P_t^{(N)}(\bm{r}_1,\ldots,\bm{r}_\ell,\ldots,\bm{r}_N)
\nonumber\displaybreak[0]\\
&\qquad=
P_t^{(N-1)}(\bm{r}_1,\ldots,\bm{r}_{\ell-1},\bm{r}_{\ell+1},\ldots,\bm{r}_N).
\end{align}

Given a single configuration of the $N$ atoms $\{\bm{r}_1,\ldots,\bm{r}_N\}$, the snapshot density profile of the cloud is constructed by
\begin{equation}
\rho(\bm{r})=\frac{1}{N}\sum_{i=1}^Ng(\bm{r}-\bm{r}_i),
\label{eqn:rho}
\end{equation}
where $g(\bm{r})$ is a function sharply peaked around $\bm{r}=0$ with unit volume $\int d^3\bm{r}\,g(\bm{r})=1$, characterizing the resolution of the photo, and the density profile $\rho(\bm{r})$ is normalized to unity:
\begin{equation}
\int d^3\bm{r}\,\rho(\bm{r})=1.
\end{equation}
Notice that the positions of the $N$ atoms, $\{\bm{r}_1,\ldots,\bm{r}_N\}$, differ from run to run, and the density profile $\rho(\bm{r})$ changes from snapshot to snapshot. 
The average profile over all possible configurations of the $N$ atoms (over all snapshots) is given by
\begin{align}
\overline{\rho(\bm{r})}
&=\int
\prod_{\ell=1}^N d^3\bm{r}_\ell\,
P_t^{(N)}(\bm{r}_1,\ldots,\bm{r}_N)\rho(\bm{r})
\nonumber\displaybreak[0]\\
&=\int d^3\bm{r}'\,g(\bm{r}-\bm{r}')P_t^{(1)}(\bm{r}'),
\label{eqn:AverageProfileP1}
\end{align}
which gives the single-particle probability distribution $P_t^{(1)}(\bm{r})$, convoluted with the resolution function $g(\bm{r})$.

When two \textit{independent} Bose gases are overlapping, no interference fringes are observed in the single-particle distribution $P_t^{(1)}(\bm{r})$, which represents the image obtained by accumulating and superposing many snapshots.
The result of the ``independence'' in the ordinary sense is that normal (Young-type) interference is absent, as is clear from (\ref{eqn:AverageProfileP1}).
An interference pattern, however, would be found on each snapshot, due to higher-order correlations \cite{ref:JavanainenYoo}.

We define observables that can characterize interference, i.e., quantities that measure whether interference is observed or not starting from some initial state.
If fringes are present on a snapshot, we expect the density deviation
\begin{equation}
\delta\rho(\bm{r})=\rho(\bm{r})-\overline{\rho(\bm{r})}
\end{equation}
to oscillate, giving rise to spikes in its Fourier transform
\begin{equation}
\delta\tilde{\rho}(\bm{k})=\int d^3\bm{r}\,\delta\rho(\bm{r}) e^{-i\bm{k}\cdot\bm{r}}.
\end{equation}
A spike at $\bm{k}_f$ in this quantity corresponds to a spatial oscillation with fringe spacing ${2\pi}/{k_f}$.

Notice here that the phase (spatial offset) of the interference pattern varies randomly from snapshot to snapshot.
This is actually unavoidable, in order to be consistent with the independence of the two gases: this random shift smears out the fringes in the average profile $\overline{\rho(\bm{r})}$, i.e., in the single-particle distribution $P_t^{(1)}(\bm{r})$, and the ``independence'' is recovered.
In order to discard this random phase, we look at the square modulus of the Fourier spectrum, $|\delta\tilde{\rho}(\bm{k})|^2$. 
If sinusoidal patterns with a definite fringe spacing (with their random spatial offsets discarded) are \textit{typical} among all possible snapshot profiles and are found on \textit{almost all} snapshots, the spikes in the spectrum $|\delta\tilde{\rho}(\bm{k})|^2$ would remain even in its average over all possible realizations of $\{\bm{r}_1,\ldots,\bm{r}_N\}$,
\begin{equation}
S_t(\bm{k})
=\overline{|\delta\tilde{\rho}(\bm{k})|^2}
=\overline{|\tilde{\rho}(\bm{k})|^2}
-\left|
\overline{\tilde{\rho}(\bm{k})}
\right|^2.
\label{eqn:S}
\end{equation}
The \textit{typicality} is characterized by the variance, or more generally, by the covariance
\begin{equation}
C_t(\bm{k},\bm{k}')
=\overline{|\delta\tilde{\rho}(\bm{k})|^2|\delta\tilde{\rho}(\bm{k}')|^2}
-\overline{|\delta\tilde{\rho}(\bm{k})|^2}
\cdot\overline{|\delta\tilde{\rho}(\bm{k}')|^2}.
\label{eqn:C}
\end{equation}
If the average spectrum $S_t(\bm{k})$ exhibits a nontrivial spike with a vanishingly small covariance $C_t(\bm{k},\bm{k}')$, the sinusoidal pattern corresponding to the spike is expected to be observed on every snapshot.

By noting $\tilde{\rho}(\bm{k})=\tilde{g}(\bm{k})\sum_{i=1}^N e^{-i\bm{k}\cdot\bm{r}_i}/N$, one realizes that these quantities are controlled by few-particle distribution functions.
Indeed,
\begin{align}
&\overline{|\tilde{\rho}(\bm{k})|^2}
=|\tilde{g}(\bm{k})|^2
\left(
\frac{N-1}{N}I_t^{(2)}(\bm{k})
+\frac{1}{N}
\right),
\displaybreak[0]\\
&\overline{|\tilde{\rho}(\bm{k})|^2|\tilde{\rho}(\bm{k}')|^2}
\nonumber\\
&\quad
=|\tilde{g}(\bm{k})|^2|\tilde{g}(\bm{k}')|^2
\left(
\frac{(N-1)!}{N^3(N-4)!}I_t^{(4)}(\bm{k},\bm{k}')
+O\!\left(\tfrac{1}{N}\right)
\right),
\end{align}
where
\begin{align}
I_t^{(2)}(\bm{k})
=\int d^3\bm{r}_1\, d^3\bm{r}_2\,
&P_t^{(2)}(\bm{r}_1,\bm{r}_2)
 e^{ i\bm{k}\cdot(\bm{r}_1-\bm{r}_2)},
\displaybreak[0]\\
I_t^{(4)}(\bm{k},\bm{k}')
=
\int
\prod_{\ell=1}^4 d^3\bm{r}_\ell\,
&P_t^{(4)}(\bm{r}_1,\bm{r}_2,\bm{r}_3,\bm{r}_4)
\nonumber\\[-2truemm]
\times{}&
 e^{ i\bm{k}\cdot(\bm{r}_1-\bm{r}_2)+ i\bm{k}'\cdot(\bm{r}_3-\bm{r}_4)}.
\end{align}
Namely, the average fringe contrast of the $N$ particles is essentially ruled by the two-particle distribution $P_t^{(2)}$, while its fluctuation by $P_t^{(4)}$. We do not need to compute the $N$-particle distribution function $P_t^{(N)}$ in practice to discuss the average fringe spectrum and the fluctuation.
We set $g(\bm{r})=\delta^3(\bm{r})$ henceforth, which does not spoil the following arguments.

The most important feature in the two-particle probability distribution is the Hanbury Brown and Twiss (HBT) effect \cite{ref:HBT-Texts}. 
Due to the bosonic nature of the atoms, the wave function has to be symmetric under the exchange of particles. 
For instance, when two atoms with opposite momenta $k$ and $-k$ overlap, $P^{(2)}$ oscillates with a period $2\pi/k$. In fact, in the setup discussed by Javanainen and Yoo \cite{ref:JavanainenYoo}, the initial state of the clouds of bosons is formed by plane waves and $P^{(2)}(x_1,x_2)= [1+\cos k(x_1-x_2)]/2$. They showed that this HBT correlation is sufficient to describe the appearance of interference between the clouds, even if the clouds are independent. Inserting this two-particle probability distribution in the above equations reproduces analytically their numerical result.

In general, the fluctuation of the snapshot profiles $\rho(\bm{r})$ is fully characterized by the generating functional
\begin{align}
Z_t[\Phi]
&=\overline{ e^{i\int d^3\bm{r}\,\Phi(\bm{r})\rho(\bm{r})}}.
\label{eqn:DefZ}
\intertext{When $N\gg1$, it is cast into (Appendix \ref{app:GeneFunc})}
Z_t[\Phi]
&\simeq
\langle
{:}
 e^{\frac{i}{N}\int d^3\bm{r}\,
\Phi(\bm{r})\hat{\psi}^\dag(\bm{r})\hat{\psi}(\bm{r})}
{:}
\rangle_t,
\label{eqn:Zfield}
\end{align}
where ${:}\cdots{:}$ denotes normal ordering.
These are our tools for the following argument, which are essentially the same as the ones employed in \cite{ref:PolkovnikovEPL,ref:Boston,ref:PatrickRathZwerger}.

Let us look at a Gaussian state, characterized by a Gaussian characteristic functional
\begin{equation}
\langle
\hat{W}[J,J^*]
\rangle_t
=e^{2i\sqrt{N}\Re\bracket{\alpha_t}{J}} e^{-N\bra{J}\hat{\mathcal{F}}_t\ket{J}},
\label{eqn:W}
\end{equation}
where 
\begin{equation}
\hat{W}[J,J^*]
=e^{i\int d^3\bm{r}\,J(\bm{r})\hat{\psi}^\dag(\bm{r})}
 e^{i\int d^3\bm{r}\,J^*(\bm{r})\hat{\psi}(\bm{r})}
\end{equation}
is the normally-ordered Weyl operator, and $\bracket{\alpha_t}{J}=\int d^3\bm{r}\,\alpha_t^*(\bm{r})J(\bm{r})
$, 
$\bra{J}\hat{\mathcal{F}}_t\ket{J}=\int d^3\bm{r}\, d^3\bm{r}'\,
J^*(\bm{r})\mathcal{F}_t(\bm{r},\bm{r}')\times J(\bm{r}')
$.
In this case, the generating functional for the density profile, $Z_t[\Phi]$, is given by (Appendix \ref{app:GeneFunc})
\begin{align}
Z_t[\Phi]
&=
 e^{-\frac{i}{N}\int d^3\bm{r}\,\frac{\delta}{\delta J(\bm{r})}\Phi(\bm{r})\frac{\delta}{\delta J^*(\bm{r})}}
\langle\hat{W}[J,J^*]\rangle_t
\biggr|_{J,J^*=0}
\nonumber\displaybreak[0]\\
&=\frac{e^{i\bra{\alpha_t}\frac{1}{\Phi^{-1}(\hat{\bm{r}})- i\hat{\mathcal{F}}_t}\ket{\alpha_t}}}{\Det[1- i\Phi(\hat{\bm{r}})\hat{\mathcal{F}}_t]}
\nonumber\displaybreak[0]\\
&=
e^{\Tr\sum\limits_{n=1}^\infty(\ket{\alpha_t}\bra{\alpha_t}+\hat{\mathcal{F}}_t/n) i\Phi(\hat{\bm{r}})[\hat{\mathcal{F}}_t i\Phi(\hat{\bm{r}})]^{n-1}},
\label{eqn:GaussG}
\end{align}
where we have introduced an abstract notation by
\begin{gather}
\alpha_t(\bm{r})=\bracket{\bm{r}}{\alpha_t},\quad
\mathcal{F}_t(\bm{r},\bm{r}')
=\bra{\bm{r}}\hat{\mathcal{F}}_t\ket{\bm{r}'},
\end{gather}
and
\begin{gather}
\hat{\bm{r}}\ket{\bm{r}}=\bm{r}\ket{\bm{r}},\\
\bracket{\bm{r}}{\bm{r}'}=\delta^3(\bm{r}-\bm{r}'),\quad
\int d^3\bm{r}\,\ket{\bm{r}}\bra{\bm{r}}=\openone.
\end{gather}

\section{Canonical Ensemble of Ideal Bosonic Atoms in a 3D Harmonic Trap}
\label{sec:SingleGas}
In order to discuss the interference of two independent ideal Bose gases released from two separate harmonic traps in 3D, we need to describe the state of a bosonic system with a fixed number of atoms.
Let us first consider a single gas held in a trap and see how it is characterized by a characteristic functional $\langle\hat{W}[J,J^*]\rangle$.
The gas consists of a \textit{fixed number}, $N$, of bosonic atoms and is kept in a harmonic trap at a finite temperature $T$.

We assume that the 3D harmonic trap is isotropic and is characterized by a trapping frequency $\omega$ (generalization to an anisotropic potential is straightforward).
Let $\ket{\varphi_{\bm{n}}}$ [$\bm{n}=(n_x,n_y,n_z)$ with $n_{x,y,z}=0,1,2,\ldots$] denote the energy eigenstates of this harmonic trap.
These eigenstates form a complete orthonormal set of bases,
\begin{equation}
\bracket{\varphi_{\bm{n}}}{\varphi_{\bm{n}'}}=\delta_{\bm{n}\bm{n}'},\qquad
\sum_{\bm{n}}\ket{\varphi_{\bm{n}}}\bra{\varphi_{\bm{n}}}=\openone,
\end{equation}
and the field operator $\hat{\psi}(\bm{r})$ is expanded as
\begin{equation}
\hat{\psi}(\bm{r})
=\sum_{\bm{n}}\hat{a}_{\bm{n}}\varphi_{\bm{n}}(\bm{r}),
\end{equation}
with $\hat{a}_{\bm{n}}$ satisfying the canonical commutation relations
\begin{equation}
[\hat{a}_{\bm{n}},\hat{a}_{\bm{n}'}^\dag]=\delta_{\bm{n}\bm{n}'},\quad
\text{etc.}
\end{equation}
The Hamiltonian of the system reads
\begin{equation}
\hat{H}=\sum_{\bm{n}}\varepsilon_{\bm{n}}
\hat{a}_{\bm{n}}^\dag
\hat{a}_{\bm{n}},\quad
\varepsilon_{\bm{n}}=\sum_{i=x,y,z}\hbar\omega n_i,
\end{equation}
and the number operator
\begin{equation}
\hat{N}=\sum_{\bm{n}}
\hat{a}_{\bm{n}}^\dag
\hat{a}_{\bm{n}}.
\end{equation}

When the gas is cooled below the critical temperature $T_c$, the ground state $\ket{\varphi_0}$ is occupied by a macroscopic number of atoms.
In the regime
\begin{equation}
\hbar\omega/k_BT\ll1\quad \text{with}\quad (\hbar\omega/k_BT)^3N\quad\text{finite}, 
\label{eqn:Regime}
\end{equation}
which is relevant in the actual experiments, the condensation fraction $\lambda$ is well approximated by (\cite{ref:BEC-Stringari,ref:Ketterle-BECHarmonic-PRA}
 and Appendices \ref{app:Canonical} and \ref{app:Purity})
 \begin{equation}
\lambda
\simeq\begin{cases}
\medskip
\displaystyle
1-\left(\frac{T}{T_c}\right)^3&(T<T_c),\\
0&(T\ge T_c),
\end{cases}
\label{eqn:Fraction}
\end{equation}
with the critical temperature given by
\begin{equation}
T_c=\frac{\hbar\omega}{k_B}\left(\frac{N}{\zeta(3)}\right)^{1/3},
\label{eqn:Tc}
\end{equation}
where $k_B$ is the Boltzmann constant and $\zeta(z)$ the Riemann zeta function.

Since the number $N$ of atoms in the gas is fixed, the gas should be described by a  \textit{canonical ensemble}, which is characterized by the characteristic functional defined by
\begin{equation}
\langle\hat{W}[J,J^*]\rangle_N
=\frac{\Tr\{\hat{W}[J,J^*]\hat{P}_Ne^{-\beta\hat{H}}\}}{\Tr\{\hat{P}_Ne^{-\beta\hat{H}}\}},
\label{eqn:CharFuncC}
\end{equation}
where $\hat{P}_N$ is the projection operator onto the $N$-particle sector
and $\beta=1/k_BT$ the inverse temperature.
In the regime (\ref{eqn:Regime}), it is estimated to be (\cite{ref:BECCanonical} and Appendix \ref{app:Canonical}) 
\begin{equation}
\langle\hat{W}[J,J^*]\rangle_N
\simeq 
\begin{cases}
\medskip
\displaystyle
J_0(2\sqrt{N}|\bracket{\alpha}{J}|)
 e^{-N\bra{J}\hat{\mathcal{F}}'\ket{J}}\ \ %
(T<T_c),\\
 e^{-N\bra{J}\hat{\mathcal{F}}\ket{J}}\hfill
(T\ge T_c),
\end{cases}
\label{eqn:CharFuncCBessel}
\end{equation}
where
\begin{gather}
\ket{\alpha}=\sqrt{\lambda}\ket{\varphi_0},\quad
\hat{\mathcal{F}}'
=\frac{1}{N}\sum_{\bm{n}\neq0}\ket{\varphi_{\bm{n}}}f(\varepsilon_{\bm{n}})\bra{\varphi_{\bm{n}}},
\label{eqn:alpha0F0'}
\displaybreak[0]\\
\hat{\mathcal{F}}
=\begin{cases}
\medskip
\ket{\alpha}\bra{\alpha}+\hat{\mathcal{F}}'&(T<T_c),\\
\displaystyle
\frac{1}{N}\sum_{\bm{n}}\ket{\varphi_{\bm{n}}}f(\varepsilon_{\bm{n}})\bra{\varphi_{\bm{n}}}&(T\ge T_c),
\end{cases}
\label{eqn:F0}
\end{gather}
with the Bose distribution function 
\begin{equation}
f(\varepsilon)=\frac{1}{e^{\beta(\varepsilon-\mu)}-1}.
\label{eqn:Bose}
\end{equation}
The chemical potential $\mu\,(\le0)$ is fixed by the condition
\begin{equation}
\sum_{\bm{n}}f(\varepsilon_{\bm{n}})=N
\label{eqn:CondMu}
\end{equation}
and is vanishingly small for $T<T_c$.
By noting a formula for the Bessel function
\begin{equation}
J_0(x)=\int_0^{2\pi}\frac{d\theta}{2\pi}\,e^{i x\cos\theta},
\end{equation}
the characteristic functional (\ref{eqn:CharFuncCBessel}) for $T<T_c$ is equivalently expressed as
\begin{align}
&\langle\hat{W}[J,J^*]\rangle_N
=\int_0^{2\pi}\frac{d\theta}{2\pi}\,
e^{2i\sqrt{N}\Re\bracket{\alpha_\theta}{J}}
e^{-N\bra{J}\hat{\mathcal{F}}'\ket{J}}
\nonumber\\[-2truemm]
&\hspace*{65truemm}
(T<T_c),
\label{eqn:CharFuncCGauss}
\end{align}
where
\begin{equation}
\ket{\alpha_\theta}=e^{i\theta}\ket{\alpha}
\end{equation}
represents a condensate with a definite phase $\theta$, and the characteristic functional (\ref{eqn:CharFuncCBessel}) for $T<T_c$ is an incoherent mixture of the condensed states with different phases $\theta$.

Notice that the grand canonical ensemble yields (Appendix \ref{app:Canonical})
\begin{equation}
\langle\hat{W}[J,J^*]\rangle_G
=\frac{\Tr\{\hat{W}[J,J^*]e^{-\beta(\hat{H}-\mu\hat{N})}\}}{\Tr\{e^{-\beta(\hat{H}-\mu\hat{N})}\}}
=e^{-N\bra{J}\hat{\mathcal{F}}\ket{J}}
\label{eqn:CharFuncGC}
\end{equation}
with $\hat{\mathcal{F}}$ given in (\ref{eqn:F0}) and $N$ being the average number of atoms, i.e., a different characteristic functional from the one for the canonical ensemble (\ref{eqn:CharFuncCBessel}) below the critical temperature, while they coincide above.
It is possible to apply the formula (\ref{eqn:GaussG}) for both canonical (\ref{eqn:CharFuncCGauss}) and grand canonical (\ref{eqn:CharFuncGC}) ensembles to obtain the generating functionals for the density profile, $Z_t[\Phi]$.
Both yield the same average profile [the single-particle distribution; see (\ref{eqn:AverageProfileP1})]
\begin{equation}
\overline{\rho(\bm{r})}
=\mathcal{F}(\bm{r},\bm{r})
\end{equation}
over the whole temperature range, while they exhibit different fluctuations below the critical temperature $T<T_c$.

\section{Two Independent Ideal Bose Gases Released from Two Separate Harmonic Traps}
\label{sec:Z}
Next, we describe the release and free expansion of the gases and derive the state just before the measurement.
Let us consider two ideal Bose gases independently prepared in two spatially separated harmonic traps in 3D\@. 
We assume that the two harmonic traps are of the same shape, characterized by the same trapping frequency $\omega$, but spatially shifted by vectors $\pm\bm{d}/2$.
The same number of atoms are put in the traps, $N$ atoms for each, at the same temperature $T$.
No particle flow is present between the two traps.
The gases are then released by turning off the trapping potential and expand in free space.
We are going to discuss the interference between the overlapping gases.

The energy eigenstates of the right and left traps are given by shifting the eigenstates $\ket{\varphi_{\bm{n}}}$ introduced in the previous section,
\begin{equation}
\ket{\varphi_{\bm{n}}^{(\pm)}}=e^{\mp\frac{i}{\hbar}\hat{\bm{p}}\cdot\bm{d}/2}\ket{\varphi_{\bm{n}}},
\label{eqn:SpatialShift}
\end{equation}
where $\hat{\bm{p}}$ is the momentum operator of an atom, which is the generator of the spatial shifts. 
We assume that the two traps are well separated, compared with the sizes of the gases, and the overlaps between the relevant eigenfunctions of the two traps are negligible:
$\bracket{\varphi_{\bm{n}}^{(+)}}{\varphi_{\bm{n}'}^{(-)}}\simeq0$.
Under this hypothesis, they form a complete orthonormal set of bases for the present system, 
\begin{gather}
\bracket{\varphi_{\bm{n}}^{(s)}}{\varphi_{\bm{n}'}^{(s')}}=\delta_{ss'}\delta_{\bm{n}\bm{n}'},\quad
\sum_{s=\pm}\sum_{\bm{n}}\ket{\varphi_{\bm{n}}^{(s)}}\bra{\varphi_{\bm{n}}^{(s)}}
=\openone,
\end{gather}
and the field operator $\hat{\psi}(\bm{r})$ is expanded as
\begin{equation}
\hat{\psi}(\bm{r})
=\sum_{s=\pm}\sum_{\bm{n}}\hat{a}_{\bm{n}}^{(s)}\varphi_{\bm{n}}^{(s)}(\bm{r}).
\end{equation}
The annihilation operators $\hat{a}_{\bm{n}}^{(s)}$ satisfy the canonical commutation relations
\begin{equation}
[\hat{a}_{\bm{n}}^{(s)},\hat{a}_{\bm{n}'}^{(s')\dag}]=\delta_{ss'}\delta_{\bm{n}\bm{n}'},\quad
\text{etc.}
\end{equation}

The expansion of the gases in free space after the release from the traps is easily implemented.
Since the gases are ideal and noninteracting, the field operator evolves in the Heisenberg picture as
\begin{equation}
\hat{\psi}(\bm{r},t)
=\sum_{s=\pm}\sum_{\bm{n}}\hat{a}_{\bm{n}}^{(s)}\varphi_{\bm{n},t}^{(s)}(\bm{r})
\end{equation}
with $\varphi_{\bm{n},t}^{(\pm)}(\bm{r})=e^{\frac{i\hbar t}{2m}\nabla^2}\varphi_{\bm{n}}^{(\pm)}(\bm{r})$: we have only to replace
\begin{equation}
\ket{\varphi_{\bm{n}}^{(\pm)}}
\to 
\ket{\varphi_{\bm{n},t}^{(\pm)}}
=e^{-\frac{i}{\hbar}\frac{\hat{\bm{p}}^2}{2m}t}\ket{\varphi_{\bm{n}}^{(\pm)}}
\label{eqn:TimeShift}
\end{equation}
in any formulas, to introduce the time development.

The two gases in the separate traps are \textit{independent} and the state of the couple is just a \textit{product state}.
Each gas is described by the canonical ensemble with a fixed number $N$ of atoms, and the characteristic functional for the couple is given by the product of the two characteristic functionals for the individual gases, each of which is given by (\ref{eqn:CharFuncCBessel}), or equivalently (\ref{eqn:CharFuncCGauss}), but shifted in space.
Then, the generating functional for the density profile of the expanding and overlapping gases in free space after the release from the two traps is readily available: by applying the formula (\ref{eqn:GaussG}) to the product state under the time evolution and by performing the integrations over the phases of the two condensates, we get
\begin{equation}
Z_t^N[\Phi]
=\begin{cases}
\displaystyle
\frac{e^{\frac{i}{2}\sum\limits_{s=\pm}\bra{\alpha_t^{(s)}}\frac{1}{\Phi^{-1}(\hat{\bm{r}})-i\hat{\mathcal{G}}_t'}\ket{\alpha_t^{(s)}}}}{\Det[1- i\Phi(\hat{\bm{r}})\hat{\mathcal{G}}_t']}
\\
\displaystyle
\quad
{}\times
J_0\biggl(\biggl|
\bra{\alpha_t^{(+)}}\frac{1}{\Phi^{-1}(\hat{\bm{r}})-i\hat{\mathcal{G}}_t'}\ket{\alpha_t^{(-)}}
\biggr|\biggr)\\
\hfill(T<T_c),\\
\displaystyle
\frac{1}{\Det[1- i\Phi(\hat{\bm{r}})\hat{\mathcal{G}}_t]}
\qquad\qquad\qquad\quad
(T\ge T_c),
\end{cases}
\label{eqn:ZcoupleC}
\end{equation}
where
\begin{equation}
\hat{\mathcal{G}}_t
=\frac{1}{2}(\hat{\mathcal{F}}_t^{(+)}+\hat{\mathcal{F}}_t^{(-)}),
\quad
\hat{\mathcal{G}}_t'
=\frac{1}{2}(\hat{\mathcal{F}}_t^{(+)\prime}+\hat{\mathcal{F}}_t^{(-)\prime}),
\end{equation}
and $\ket{\alpha_t^{(\pm)}}$, $\hat{\mathcal{F}}_t^{(\pm)\prime}$, $\hat{\mathcal{F}}_t^{(\pm)}$ are defined by shifting (\ref{eqn:alpha0F0'})--(\ref{eqn:F0}) in space and time by the unitary transformations (\ref{eqn:SpatialShift}) and (\ref{eqn:TimeShift}), i.e.,
\begin{gather}
\ket{\alpha_t^{(\pm)}}
=e^{-\frac{i}{\hbar}\frac{\hat{\bm{p}}^2}{2m}t}
e^{\mp\frac{i}{\hbar}\hat{\bm{p}}\cdot\bm{d}/2}
\ket{\alpha},\\
\hat{\mathcal{F}}_t^{(\pm)}
=e^{-\frac{i}{\hbar}\frac{\hat{\bm{p}}^2}{2m}t}
e^{\mp\frac{i}{\hbar}\hat{\bm{p}}\cdot\bm{d}/2}
\hat{\mathcal{F}}
e^{\pm\frac{i}{\hbar}\hat{\bm{p}}\cdot\bm{d}/2}
e^{\frac{i}{\hbar}\frac{\hat{\bm{p}}^2}{2m}t},
\ \text{etc.}
\end{gather}

The grand canonical ensembles (\ref{eqn:CharFuncGC}), on the other hand, yield
\begin{equation}
Z_t^G[\Phi]
=\frac{1}{\Det[1- i\Phi(\hat{\bm{r}})\hat{\mathcal{G}}_t]}
\label{eqn:ZcoupleGC}
\end{equation}
over the whole temperature range.
It coincides with the one for the canonical ensembles above the critical temperature $T\ge T_c$, while they are different below $T<T_c$.

\section{Interference and Fluctuation}
\label{sec:InterferenceFluctuation}
We are now ready to discuss the interference between the two gases released from the two harmonic traps, applying the tools introduced in Sec.\ \ref{sec:Tools} to the state obtained in Sec.\ \ref{sec:Z}.

We first concentrate on the generating functional (\ref{eqn:ZcoupleC}) for the canonical ensembles, which is rewritten as
\begin{align}
Z_t^N[\Phi]
=\begin{cases}
\displaystyle
 e^{\Tr\sum\limits_{n=1}^\infty\Bigl(
\sum\limits_{s=\pm}\frac{1}{2}\ket{\alpha_t^{(s)}}\bra{\alpha_t^{(s)}}
+\frac{1}{n}\hat{\mathcal{G}}_t'
\Bigr)\, 
i\Phi(\hat{\bm{r}})[\hat{\mathcal{G}}_t' i\Phi(\hat{\bm{r}})]^{n-1}}
\\
\displaystyle
{}\times
J_0\biggl(\biggl|
\sum\limits_{n=1}^\infty
\bra{\alpha_t^{(+)}}
i\Phi(\hat{\bm{r}})[\hat{\mathcal{G}}_t' i\Phi(\hat{\bm{r}})]^{n-1}
\ket{\alpha_t^{(-)}}
\biggr|\biggr)
\\
\hfill
(T< T_c),\\
\displaystyle
 e^{\Tr\sum\limits_{n=1}^\infty
\frac{1}{n}[\hat{\mathcal{G}}_t i\Phi(\hat{\bm{r}})]^n}
\hfill
(T\ge T_c).
\end{cases}
\label{eqn:ZcoupleCexp}
\end{align}
By noting the series expansion of the Bessel function
\begin{equation}
J_0(x)=1-\frac{1}{4}x^2+\frac{1}{64}x^4+\cdots,
\end{equation}
the average profile is immediately obtained as
\begin{equation}
\overline{\rho(\bm{r})}
=\mathcal{G}_t(\bm{r},\bm{r})
=\frac{1}{2}\,\Bigl(\hat{\mathcal{F}}_t^{(+)}(\bm{r},\bm{r})+\hat{\mathcal{F}}_t^{(-)}(\bm{r},\bm{r})\Bigr).
\label{eqn:RhoBar}
\end{equation}
This is just the sum of the density profiles of the two clouds, and no interference is observed in this quantity.
This is the ``independence'' of the two gases.
However, interference fringes are found on each snapshot.
The average spectrum defined in (\ref{eqn:S}) is given in the present case by
\begin{align}
S_t(\bm{k})
={}&\Tr\{\hat{\mathcal{G}}_te^{i\bm{k}\cdot\hat{\bm{r}}}\hat{\mathcal{G}}_te^{-i\bm{k}\cdot\hat{\bm{r}}}\}
\nonumber\\
&{}-\frac{1}{4}\sum_{s=\pm}|\bra{\alpha_t^{(s)}}e^{i\bm{k}\cdot\hat{\bm{r}}}\ket{\alpha_t^{(s)}}|^2,
\label{eqn:SC}
\end{align}
both below and above the critical temperature.
Since the generic formula for the covariance (\ref{eqn:C}) is too complicated, we just give its expressions for two extreme cases: at zero temperature (pure condensates), 
\begin{align}
&C_t(\bm{k},\bm{k}')
\nonumber\\
&\quad
=\frac{1}{8}\Re\Bigl(
\bra{\alpha_t^{(+)}}e^{i\bm{k}\cdot\hat{\bm{r}}}\ket{\alpha_t^{(-)}}
\bra{\alpha_t^{(+)}}e^{-i\bm{k}\cdot\hat{\bm{r}}}\ket{\alpha_t^{(-)}}
\nonumber\\
&\quad\qquad\qquad
{}\times
\bra{\alpha_t^{(-)}}e^{i\bm{k}'\cdot\hat{\bm{r}}}\ket{\alpha_t^{(+)}}
\bra{\alpha_t^{(-)}}e^{-i\bm{k}'\cdot\hat{\bm{r}}}\ket{\alpha_t^{(+)}}
\Bigr)
\nonumber\\
&\hspace*{70truemm}
(T=0),
\label{eqn:CC0}
\end{align}
and above the critical temperature (in the absence of condensates),
\begin{align}
&C_t(\bm{k},\bm{k}')
\nonumber\\
&\quad
=\sum_{\bm{q}=\pm\bm{k}}
\sum_{\bm{q}'=\pm\bm{k}'}
\Tr\{\hat{\mathcal{G}}_t e^{ i\bm{q}\cdot\hat{\bm{r}}}\hat{\mathcal{G}}_t e^{- i\bm{q}\cdot\hat{\bm{r}}}
\hat{\mathcal{G}}_t e^{ i\bm{q}'\cdot\hat{\bm{r}}}\hat{\mathcal{G}}_t e^{- i\bm{q}'\cdot\hat{\bm{r}}}\}
\nonumber\\
&\quad\quad
{}+
\sum_{\bm{q}'=\pm\bm{k}'}
\Tr\{
\hat{\mathcal{G}}_t e^{ i\bm{k}\cdot\hat{\bm{r}}}\hat{\mathcal{G}}_t e^{- i\bm{q}'\cdot\hat{\bm{r}}}\hat{\mathcal{G}}_t e^{- i\bm{k}\cdot\hat{\bm{r}}}
\hat{\mathcal{G}}_t e^{ i\bm{q}'\cdot\hat{\bm{r}}}
\}
\nonumber\\
&\quad\quad
{}+
\sum_{\bm{q}'=\pm\bm{k}'}
|{\Tr\{\hat{\mathcal{G}}_t e^{ i\bm{k}\cdot\hat{\bm{r}}}\hat{\mathcal{G}}_t e^{- i\bm{q}'\cdot\hat{\bm{r}}}\}}|^2
\qquad\qquad
(T\ge T_c).
\label{eqn:CCT}
\end{align}

\subsection{At Zero Temperature $T=0$}
Let us look at the zero-temperature case in detail.
In this case, the average profile (\ref{eqn:RhoBar}) is given by 
\begin{equation}
\overline{\rho(\bm{r})}
=\frac{1}{2}\,\Bigl(
|\alpha_t^{(+)}(\bm{r})|^2
+|\alpha_t^{(-)}(\bm{r})|^2
\Bigr),
\label{eqn:RhoBarZero}
\end{equation}
the average spectrum (\ref{eqn:SC}) is reduced to 
\begin{equation}
S_t(\bm{k})
=\frac{1}{4}\,\Bigl(
|\chi_t(\bm{k})|^2
+|\chi_t(-\bm{k})|^2
\Bigr),
\label{eqn:SC0}
\end{equation}
and the covariance (\ref{eqn:CC0}) 
\begin{equation}
C_t(\bm{k},\bm{k}')
=\frac{1}{8}\Re\Bigl(
\chi_t^*(\bm{k})
\chi_t^*(-\bm{k})
\chi_t(-\bm{k}')
\chi_t(\bm{k}')
\Bigr),
\label{eqn:CC0chi}
\end{equation}
both given in terms of 
\begin{equation}
\chi_t(\bm{k})
=\bra{\alpha_t^{(+)}}e^{-i\bm{k}\cdot\hat{\bm{r}}}\ket{\alpha_t^{(-)}},
\end{equation}
which is the Fourier transform of the interference term $\alpha_t^{(+)*}(\bm{r})\alpha_t^{(-)}(\bm{r})$ between the two condensate wave functions $\alpha_t^{(\pm)}(\bm{r})$.
For the present harmonic traps, they read
\begin{equation}
\alpha_t^{(\pm)}(\bm{r})
=\left(\frac{m\omega}{\pi\hbar(1+i\omega t)^2}\right)^{3/4}e^{-m\omega(\bm{r}\mp\bm{d}/2)^2/2\hbar(1+i\omega t)}
\end{equation}
and 
\begin{equation}
\chi_t(\bm{k})
=e^{-\hbar[\bm{k}^2+(\bm{k}\omega t-m\omega \bm{d}/\hbar)^2]/4m\omega}.
\label{eqn:ChiGauss}
\end{equation}

\begin{figure}
\includegraphics{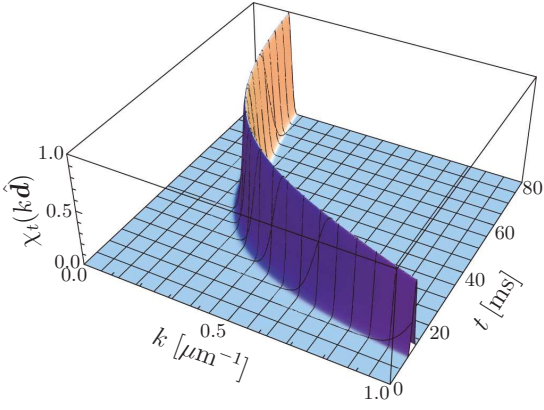}
\caption{(Color online) The time evolution of the interference spectrum $\chi_t(\bm{k})$ in  (\ref{eqn:ChiGauss}) between two pure condensates at zero temperature $T=0$.
The two condensates, each containing $N=5\times10^6$ Na atoms, are released from two harmonic traps of trapping frequency $\omega=1.6\,\text{kHz}$, separated by a distance $d=30\,\mu\text{m}$.
The condensates expand, overlap, and exhibit an interference pattern on each snapshot. The critical temperature of the gas trapped in this harmonic potential is estimated by (\ref{eqn:Tc}) to be $T_c=2.0\,\mu\text{K}$\@.}
\label{fig:chi}
\end{figure}
The time evolution of $\chi_t(\bm{k})$ is shown in Fig.\ \ref{fig:chi}, for an experimentally realistic set of parameters.
Sharp peaks grow in the average spectrum $S_t(\bm{k})$ given in (\ref{eqn:SC0}) at
\begin{equation}
\bm{k}=\pm\bm{k}_f,\quad
\bm{k}_f=\frac{m\bm{d}}{\hbar t}.
\label{eqn:kf}
\end{equation}
The peaks become sharper and higher as time goes on.
The covariance $C_t(\bm{k},\bm{k}')$ in (\ref{eqn:CC0chi}), on the other hand, is vanishingly small for any $(\bm{k},\bm{k}')$, since the peaks of $\chi_t(\bm{k})$ and $\chi_t(-\bm{k})$ are well separated \cite{ref:PolkovnikovEPL}.
This means that there is \textit{no fluctuation} in $|\delta\tilde{\rho}(\bm{k})|^2$ around the average $S_t(\bm{k})$ in (\ref{eqn:SC0}), and \textit{each single snapshot} exhibits a profile 
\begin{equation}
\delta\tilde{\rho}(\bm{k})
\simeq\frac{1}{2}\chi_t(\bm{k})e^{-i\delta_t(\bm{k})}
+\frac{1}{2}\chi_t(-\bm{k})e^{i\delta_t(-\bm{k})},
\label{eqn:SnapshotSpectrum}
\end{equation}
with an unknown phase $\delta_t(\bm{k})$.
Note that there is essentially no overlap between $\chi_t(\bm{k})$ and $\chi_t(-\bm{k})$, and that $\delta\rho(\bm{r})$ is a real function, namely, $\delta\tilde{\rho}(\bm{k})=\delta\tilde{\rho}^*(-\bm{k})$.
In addition, the phase $\delta_t(\bm{k})$ should change randomly from snapshot to snapshot to be consistent with $\overline{\delta\rho(\bm{k})}=0$ [see (\ref{eqn:RhoBarZero})].
By replacing $\delta_t(\bm{k})\to\delta_f=\delta_t(\bm{k}_f)$ since the spectrum is very sharp at $\bm{k}=\pm\bm{k}_f$, and by performing the inverse Fourier transform of (\ref{eqn:SnapshotSpectrum}), we get a snapshot profile
\begin{align}
\rho(\bm{r})
&=\overline{\rho(\bm{r})}+\delta\rho(\bm{r})
\nonumber
\\
&\simeq\frac{1}{2}\left|
\alpha_t^{(+)}(\bm{r})
e^{i\delta_f/2}
+\alpha_t^{(-)}(\bm{r})
e^{-i\delta_f/2}
\right|^2,
\label{eqn:RhoSnapshot0}
\end{align}
which exhibits an interference pattern with \textit{perfect visibility}, with a spatial offset $\delta_f$.
Note that the visibility is essentially ruled by the height of the spectrum $S_t(\pm\bm{k}_f)$ with its maximum $1/4$.

We stress that $\rho(\bm{r})$ obtained in (\ref{eqn:RhoSnapshot0}) is a snapshot profile and not a quantum-mechanical average.
We started with a fixed number $N$ of atoms for each gas, and have kept the $\text{U}(1)$ symmetry of the system during the whole calculation: the symmetry breaking is not assumed.
In fact, the characteristic functional (\ref{eqn:CharFuncCBessel}) yields
\begin{equation}
\langle\hat{\psi}(\bm{r})\rangle=0.
\end{equation}
Without definite relative phase between the gases, interference would not be expected in the standard way we understand the Young-type interference.
However, a sinusoidal pattern with a definite fringe spacing $\lambda_f=2\pi/k_f=ht/md$ with a good visibility is certainly observed on every snapshot, as a result of the effects of the higher-order correlations.
Moreover, the vanishing covariance allows us to reconstruct the snapshot density profile as (\ref{eqn:RhoSnapshot0}).
These are fully consistent with a series of the previous studies \cite{ref:JavanainenYoo,ref:PolkovnikovEPL,ref:MI,ref:MI-int,ref:MI-Rev,ref:Boston,ref:PatrickRathZwerger}.

\subsection{Interference and Condensation}
\label{sec:IntCond}
For a generic finite temperature $T$, it is possible to obtain asymptotic but explicit and concise formulas for the average spectrum $S_t(\bm{k})$ and the covariance $C_t(\bm{k},\bm{k}')$ for large $t$.
Let us first observe the asymptotic behavior of the wave function $\varphi_{\bm{n},t}^{(\pm)}(\bm{r})$. 
By the method of steepest descent (stationary-phase approximation), we get \cite{ref:PolkovnikovEPL,ref:ProbeCorrelationAltman,ref:Boston}
\begin{widetext}
\begin{align}
\varphi_{\bm{n},t}^{(\pm)}(\bm{r})
=\int d^3\bm{k}\,e^{-i\hbar\bm{k}^2t/2m}e^{i\bm{k}\cdot(\bm{r}\mp\bm{d}/2)}
\tilde{\varphi}_{\bm{n}}(\bm{k})
\sim\left(
\frac{m}{i\hbar t}
\right)^{3/2}
e^{im(\bm{r}\mp\bm{d}/2)^2/2\hbar t}
\tilde{\varphi}_{\bm{n}}\!\left(
\frac{m}{\hbar t}\left(\bm{r}\mp\frac{\bm{d}}{2}\right)
\right).
\end{align}
The interference terms are then estimated to be
\begin{align}
\bra{\varphi_{\bm{n}',t}^{(\pm)}}e^{-i\bm{k}\cdot\hat{\bm{r}}}\ket{\varphi_{\bm{n},t}^{(\mp)}}
&\sim\left(
\frac{m}{\hbar t}
\right)^3\int d^3\bm{r}\,
\tilde{\varphi}_{\bm{n}'}^*\!\left(
\frac{m}{\hbar t}\left(\bm{r}\mp\frac{\bm{d}}{2}\right)
\right)
\tilde{\varphi}_{\bm{n}}\!\left(
\frac{m}{\hbar t}\left(\bm{r}\pm\frac{\bm{d}}{2}\right)
\right)
e^{-i(\bm{k}\mp m\bm{d}/\hbar t)\cdot\bm{r}}
\nonumber\displaybreak[0]\\
&=\int d^3\bm{k}'\,
\tilde{\varphi}_{\bm{n}'}^*\!\left(
\bm{k}'\mp\frac{m\bm{d}}{2\hbar t}
\right)
\tilde{\varphi}_{\bm{n}}\!\left(
\bm{k}'\pm\frac{m\bm{d}}{2\hbar t}
\right)
e^{-i\bm{k}'\cdot\hbar t(\bm{k}\mp m\bm{d}/\hbar t)/m}
\nonumber\\
&\sim\int d^3\bm{k}'\,
\tilde{\varphi}_{\bm{n}'}^*\!\left(
\bm{k}'
\right)
\tilde{\varphi}_{\bm{n}}\!\left(
\bm{k}'
\right)
e^{-i\bm{k}'\cdot\hbar t(\bm{k}\mp\bm{k}_f)/m},
\intertext{which is sharply peaked at $\bm{k}\simeq\pm\bm{k}_f$ with $\bm{k}_f$ defined in (\ref{eqn:kf}), and similarly,}
\bra{\varphi_{\bm{n}',t}^{(\pm)}}e^{-i\bm{k}\cdot\hat{\bm{r}}}\ket{\varphi_{\bm{n},t}^{(\pm)}}
&\sim\left(
\frac{m}{\hbar t}
\right)^3\int d^3\bm{r}\,
\tilde{\varphi}_{\bm{n}'}^*\!\left(
\frac{m}{\hbar t}\left(\bm{r}\mp\frac{\bm{d}}{2}\right)
\right)
\tilde{\varphi}_{\bm{n}}\!\left(
\frac{m}{\hbar t}\left(\bm{r}\mp\frac{\bm{d}}{2}\right)
\right)
e^{-i\bm{k}\cdot\bm{r}}
\nonumber\displaybreak[0]\\
&=\int d^3\bm{k}'\,
\tilde{\varphi}_{\bm{n}'}^*(\bm{k}')
\tilde{\varphi}_{\bm{n}}(\bm{k}')
e^{-i\bm{k}'\cdot\hbar t\bm{k}/m}
e^{\mp i\bm{k}\cdot(\hbar\bm{k}_ft/2m)}
\end{align}
\end{widetext}
is sharply peaked at $\bm{k}\simeq0$.
At these peaks, 
\begin{equation}
\begin{cases}
\medskip
\displaystyle
\bra{\varphi_{\bm{n}',t}^{(\pm)}}e^{-i\bm{k}\cdot\hat{\bm{r}}}\ket{\varphi_{\bm{n},t}^{(\mp)}}
\sim\bracket{\varphi_{\bm{n}'}}{\varphi_{\bm{n}}}&(\bm{k}=\pm\bm{k}_f),
\\
\displaystyle
\bra{\varphi_{\bm{n}',t}^{(\pm)}}e^{-i\bm{k}\cdot\hat{\bm{r}}}\ket{\varphi_{\bm{n},t}^{(\pm)}}
\sim\bracket{\varphi_{\bm{n}'}}{\varphi_{\bm{n}}}&(\bm{k}=0).
\end{cases}
\end{equation}
Applying these asymptotic behaviors to (\ref{eqn:ZcoupleCexp}), we get the average profile
\begin{equation}
\begin{cases}
\medskip
\displaystyle
\overline{\tilde{\rho}(0)}
\sim1,
\\
\displaystyle
\overline{\tilde{\rho}(\bm{k}_f)}
\sim0,
\end{cases}
\label{eqn:RhoBarAsymp}
\end{equation}
the average spectrum
\begin{equation}
\begin{cases}
\medskip
\displaystyle
S_t(\bm{k}_f)
\sim\frac{1}{4}\Tr\{\hat{\mathcal{F}}^2\},
\\
\displaystyle
S_t(0)
\sim\frac{1}{2}\Tr\{\hat{\mathcal{F}}'^2\},
\end{cases}
\label{eqn:Sasymp}
\end{equation}
and the covariance
\begin{equation}
\begin{cases}
\medskip
\displaystyle
C_t(\bm{k}_f,\bm{k}_f)
\sim\frac{1}{16}\left(
2\Tr\{\hat{\mathcal{F}}^4\}
+\Tr\{\hat{\mathcal{F}}^2\}^2
-3\lambda^4
\right),\\
\medskip
\displaystyle
C_t(0,0)
\sim\frac{1}{4}\left(
3\Tr\{\hat{\mathcal{F}}'^4\}
+2\Tr\{\hat{\mathcal{F}}'^2\}^2
\right),\\
\displaystyle
C_t(\bm{k}_f,0)
\sim\frac{3}{8}\Tr\{\hat{\mathcal{F}}'^4\}
\end{cases}
\label{eqn:Casymp}
\end{equation}
for large $t$, where $\hat{\mathcal{F}}$ defined in (\ref{eqn:F0}) is the single-particle density operator of each gas, $\hat{\mathcal{F}}'\,(=\hat{\mathcal{F}}-\ket{\alpha}\bra{\alpha})$ defined in (\ref{eqn:alpha0F0'}) describes the thermal excitations, and $\lambda$ is the condensation fraction. 
The above expressions are valid over the whole range of temperature $T$.
These are the main results of this paper.

\begin{figure}
\includegraphics[width=0.4\textwidth]{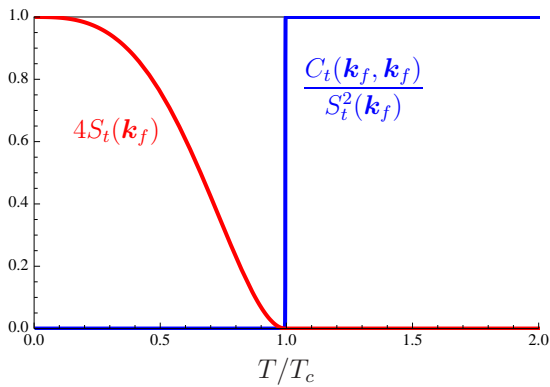}
\caption{(Color online) The average and the fluctuation of the snapshot interference spectrum, $S_t(\bm{k}_f)$ and $C_t(\bm{k}_f,\bm{k}_f)$ given in (\ref{eqn:Sasymp}) and (\ref{eqn:Casymp}), respectively, as functions of the temperature of the gases $T$. The parameters are the same as in Fig.\ \ref{fig:chi}. The relevant quantities $\lambda$, $\Tr\{\hat{\mathcal{F}}'^2\}$, and $\Tr\{\hat{\mathcal{F}}'^4\}$ are numerically evaluated without resort to the continuum limit (\ref{eqn:Scaling}).
The approximate analytical expressions (\ref{eqn:lambda}), (\ref{eqn:F2}), and (\ref{eqn:F4}) perfectly reproduce these numerical results.}
\label{fig:SC}
\end{figure}

Recall here that $\Tr\{\hat{\mathcal{F}}^2\}$ is the ``purity'' of each gas, and the average fringe spectrum $S_t(\bm{k}_f)$ in (\ref{eqn:Sasymp}) is given by the purity.
The purity is vanishingly small $\Tr\{\hat{\mathcal{F}}^2\}\sim0$ in the absence of condensate above the critical temerature $T\ge T_c$, while it becomes $\Tr\{\hat{\mathcal{F}}^2\}\sim O(1)$ as the ground state is occupied by a macroscopic number of atoms below the critical temperature $T<T_c$, approaching $\Tr\{\hat{\mathcal{F}}^2\}=1$ for pure condensation at $T=0$.
The purity is a good measure of condensation and is adopted for a criterion of BEC by Penrose and Onsager \cite{ref:CondPurity}.
The formula for the average fringe spectrum $S_t(\bm{k}_f)$ in (\ref{eqn:Sasymp}) explicitly clarifies the connection between the condensation and the interference, and the importance of the condensation for the interference.
The purity 
\begin{equation}
\Tr\{\hat{\mathcal{F}}^2\}
=\lambda^2+\Tr\{\hat{\mathcal{F}}'^2\}
\end{equation}
is different from $\lambda^2$ by $\Tr\{\hat{\mathcal{F}}'^2\}\simeq O(1/N)$ [see (\ref{eqn:F2}) in Appendix \ref{app:Purity}], and therefore, the purity is essentially given by the square of the condensation fraction $\lambda^2$ \cite{ref:PatrickRathZwerger}.
See Fig.\ \ref{fig:SC}, where the average fringe spectrum $S_t(\bm{k}_f)$ is plotted as a function of the temperature $T$.

The fluctuation of the fringe spectrum in (\ref{eqn:Casymp}) (relative to the average), on the other hand,  is estimated to be
\begin{align}
\frac{C_t(\bm{k}_f,\bm{k}_f)}{S_t^2(\bm{k}_f)}
&\sim1-\frac{
\lambda^4-2\Tr\{\hat{\mathcal{F}}'^4\}
}{(\lambda^2+\Tr\{\hat{\mathcal{F}}'^2\})^2}
\nonumber\displaybreak[0]
\\
&\simeq\begin{cases}
\medskip
O(1/N)&(T<T_c),\\
1+O(1/N)&(T\ge T_c),
\end{cases}
\label{eqn:FluctuationEst}
\end{align}
by noting that
\begin{gather}
\lambda\simeq
\begin{cases}
\medskip
O(1),\\
O(1/N),
\end{cases}
\Tr\{\hat{\mathcal{F}}'^4\}
\simeq
\begin{cases}
\medskip
O(1/N^2)&(T<T_c),\\
O(1/N^3)&(T\ge T_c),
\end{cases}
\end{gather}
and $\Tr\{\hat{\mathcal{F}}'^2\}\simeq O(1/N)$ for the whole temperature range [see (\ref{eqn:lambda}), (\ref{eqn:F2}), and (\ref{eqn:F4}) in Appendix \ref{app:Purity}].
The fluctuation is vanishingly small below the critical temperature $T<T_c$ \cite{ref:PolkovnikovEPL}, while it is nonvanishing above $T\ge T_c$.
As shown in Fig.\ \ref{fig:SC}, the fluctuation abruptly changes at the critical temperature $T_c$.
In particular, the interference spectrum does not fluctuate at any temperature below the critical temperature $T<T_c$, and in this range, the interference pattern with fringe contrast $\lambda$ is certainly observed on every snapshot.

If the gases are described by grand canonical ensembles, instead of the canonical ensembles, the statistics of the snapshot profiles are given by $Z_t^G[\Phi]$ in (\ref{eqn:ZcoupleGC}), and we end up with different conclusion from the above.
In order to switch to the grand canonical ensembles, we have only to replace $\lambda\to0$ and $\hat{\mathcal{F}}'\to\hat{\mathcal{F}}$ in (\ref{eqn:Sasymp}) and (\ref{eqn:Casymp}).
While the average fringe spectrum $S_t(\bm{k}_f)$ remains unchanged, the variance $C_t(\bm{k}_f,\bm{k}_f)$ exhibits different fluctuation with the grand canonical ensembles:
\begin{align}
\frac{C_t(\bm{k}_f,\bm{k}_f)}{S_t^2(\bm{k}_f)}
&\sim1+\frac{
2\Tr\{\hat{\mathcal{F}}^4\}
}{\Tr\{\hat{\mathcal{F}}^2\}^2}
\simeq\begin{cases}
\medskip
3+O(1/N)&(T<T_c),\\
1+O(1/N)&(T\ge T_c).
\end{cases}
\label{eqn:FluctuationEstG}
\end{align}
The fringe spectrum largely fluctuates below the critical temperature $T<T_c$, in contrast to the vanishing fluctuation with the canonical ensembles in (\ref{eqn:FluctuationEst}) and in Fig.\ \ref{fig:SC}.

The main difference between the canonical and grand canonical ensembles is the fluctuation of the total number of atoms.
In the case of canonical ensembles, it is given by $S_t(0)$ in (\ref{eqn:Sasymp}), which is vanishingly small compared to the average $\overline{\tilde{\rho}(0)}=1$ in (\ref{eqn:RhoBarAsymp}), over the whole temperature region \cite{note:Purity}. 
In the case of the grand canonical ensembles, on the other hand, it is given by
\begin{equation}
S_t(0)\sim\frac{1}{2}\Tr\{\hat{\mathcal{F}}^2\}
\simeq\begin{cases}
\medskip
\displaystyle
\frac{1}{2}\lambda^2+O(1/N)&(T<T_c),\\
O(1/N)&(T\ge T_c),
\end{cases}
\end{equation}
and the total number of atoms becomes fluctuating below the critical temperature $T<T_c$.
Although usually the canonical and grand canonical ensembles coincide in the thermodynamical limit $N\to\infty$, it is not the case in the presence of condensate.
This difference leads to the difference in the fluctuation of the fringe spectrum in (\ref{eqn:FluctuationEst}) and (\ref{eqn:FluctuationEstG}).
From a mathematical point of view, the Bessel function $J_0$ characteristic in the generating functional $Z_t^N[\Phi]$ for the canonical ensemble (\ref{eqn:ZcoupleC}) leads to the suppression of the fluctuation in the fringe spectrum below the critical temperature $T<T_c$ in (\ref{eqn:FluctuationEst}).
The canonical ensemble, in which the total number of atoms is fixed, is important for the interference pattern to be certainly observed on every snapshot below the critical temperature $T<T_c$.

\section{Summary}
\label{sec:Conclusion}
We have studied the interference of two independently prepared ideal gases of bosonic atoms, on the basis of the idea of measurement-induced interference.
The number of atoms in each gas, $N$, is fixed finite, and the $\text{U}(1)$ symmetry of the system is not broken.
Interference fringes are however observed on each snapshot, as a result of the higher-order correlations among the identical particles.
In this paper, we are interested, in particular, in the relevance of the Bose-Einstein condensation to this phenomenon \cite{ref:PatrickRathZwerger}.
Each gas with the definite number of atoms $N$ is described by a canonical ensemble trapped in a 3D harmonic potential at a finite temperature $T$ [Eq.\ (\ref{eqn:CharFuncCBessel})], and the characteristic functional $Z_t^N[\Phi]$ for the statistics of the snapshot profiles of the cloud of the overlapping gases released from two spatially separated traps is computed [Eq.\ (\ref{eqn:ZcoupleC})].
A concise formula is then obtained which clarifies the relationship between the Bose-Einstein condensation and the interference: the average fringe spectrum $S_t(\bm{k}_f)$ is given by the purity $\Tr\{\hat{\mathcal{F}}^2\}$ of each gas [Eq.\ (\ref{eqn:Sasymp})], which in turn is a good measure of condensation \cite{ref:CondPurity}. 
The fluctuation of the fringe spectrum is also analyzed [Eq.\ (\ref{eqn:Casymp})], and the fluctuation is shown to be vanishingly small below the critical temperature $T<T_c$ \cite{ref:PolkovnikovEPL} while it is nonvanishing above $T\ge T_c$ [Eq.\ (\ref{eqn:FluctuationEst}) and Fig.\ \ref{fig:SC}].
For this vanishing fluctuation, the canonical ensemble (the fact that the number of atoms in each gas is fixed) is important.

In the present paper, as well as in most of the previous works, the measurement-induced interference has been studied with ideal Bose gases.
It is an important subject to clarify the effects of the intra-atomic interaction in the gases \cite{ref:MI-int}.
It is important to keep in mind that the single-particle density operator $\hat{\mathcal{F}}$ is well defined even for an interacting gas, and even in this case the purity $\Tr\{\hat{\mathcal{F}}^2\}$ is a good measure for the degree of condensation of the gas \cite{ref:CondPurity}.
It would be tempting to apply the same reasoning as the present analysis for interacting gases and to see how the interaction affects the fringe contrast and its fluctuation. 
However, in the case of strongly interacting systems, the generating functional can be substantially different from the one considered here.
It is also an interesting problem how to deal with the interaction during the expansion of the gases in a self-consistent way, at least at the initial stages.
These issues deserve investigation.

\begin{acknowledgments}
The authors wish to thank Shuichi Tasaki for helpful discussions and encouragement. This work is supported by a Special Coordination Fund for Promoting Science and Technology, and the Grant-in-Aid for Young Scientists (B) (No.\ 21740294), both from the Ministry of Education, Culture, Sports, Science and Technology (MEXT), Japan,
by the bilateral Italian-Japanese Projects II04C1AF4E on ``Quantum Information, Computation and Communication'' of the Italian Ministry of Education, University and Research (MIUR), and
by the Joint Italian-Japanese Laboratory on ``Quantum Information and Computation'' of the Italian Ministry for Foreign Affairs (MAE).
\end{acknowledgments}

\appendix
\section{Generating Functional for Snapshot Profiles}
\label{app:GeneFunc}
In this appendix, we sketch the derivation of the formula (\ref{eqn:Zfield}) for the generating functional for the snapshot profiles $Z_t[\Phi]$ and its application to the Gaussian state characterized by the Gaussian characteristic functional $\langle\hat{W}[J,J^*]\rangle_t$ in (\ref{eqn:W}) to obtain $Z_t[\Phi]$ in (\ref{eqn:GaussG}).

The snapshot density profile $\rho(\bm{r})$ in (\ref{eqn:rho}) fluctuates from snapshot to snapshot, since the configuration of the atoms $\{\bm{r}_1,\ldots,\bm{r}_N\}$ changes from run to run according to the probability distribution $P_t^{(N)}$.
The statistics of the snapshot profiles is characterized by the generating functional $Z_t[\Phi]$ defined in (\ref{eqn:DefZ}), i.e.,
\begin{equation}
Z_t[\Phi]
=\int \prod_{\ell=1}^N d^3\bm{r}_\ell\,
P_t^{(N)}(\bm{r}_1,\ldots,\bm{r}_N)
 e^{i\int d^3\bm{r}\,\Phi(\bm{r})\rho(\bm{r})}.
\end{equation}
Setting $g(\bm{r})=\delta^3(\bm{r})$ without loss of the essence of the discussion, it is arranged in the following way to obtain the formula in (\ref{eqn:Zfield}):
\begin{widetext}
\begin{align}
Z_t[\Phi]
&=\int \prod_{\ell=1}^N d^3\bm{r}_\ell\,
P_t^{(N)}(\bm{r}_1,\ldots,\bm{r}_N)
 e^{\frac{i}{N}\sum\limits_{i=1}^N\Phi(\bm{r}_i)}
\nonumber
\displaybreak[0]
\\
&\simeq\sum_{M=0}^\infty
\frac{1}{M!}
\left(\frac{i}{N}\right)^M
\int \prod_{\ell=1}^N d^3\bm{r}_\ell\,
P_t^{(N)}(\bm{r}_1,\ldots,\bm{r}_N)
\mathop{\sum\cdots\sum}_{i_1\neq \cdots\neq i_M}\Phi(\bm{r}_{i_1})\cdots\Phi(\bm{r}_{i_M})\quad
(N\gg1)
\nonumber
\displaybreak[0]
\\
&=\sum_{M=0}^\infty
\frac{1}{M!}
\left(\frac{i}{N}\right)^M
\frac{N!}{(N-M)!}
\int \prod_{\ell=1}^M d^3\bm{r}_\ell\,
P_t^{(M)}(\bm{r}_1,\ldots,\bm{r}_M)
\Phi(\bm{r}_1)\cdots\Phi(\bm{r}_M)
\nonumber
\displaybreak[0]
\\
&=\sum_{M=0}^\infty
\frac{1}{M!}
\left(\frac{i}{N}\right)^M
\int \prod_{\ell=1}^M d^3\bm{r}_\ell\,
\langle\hat{\psi}^\dag(\bm{r}_1)\cdots\hat{\psi}^\dag(\bm{r}_M)\hat{\psi}(\bm{r}_M)\cdots\hat{\psi}(\bm{r}_1)\rangle_t
\nonumber
\displaybreak[0]
\\
&=
\langle
{:}
 e^{\frac{i}{N}\int d^3\bm{r}\,
\Phi(\bm{r})\hat{\psi}^\dag(\bm{r})\hat{\psi}(\bm{r})}
{:}
\rangle_t.
\end{align}
\end{widetext}

\subsection*{For Gaussian States}
\label{app:GeneFuncGauss}
For the Gaussian state characterized by the Gaussian characteristic functional $\langle\hat{W}[J,J^*]\rangle_t$ in (\ref{eqn:W}), the generating functional for the snapshot profiles $Z_t[\Phi]$ in (\ref{eqn:Zfield}) is computed as
\begin{align}
&e^{-i\int d^3\bm{r}\,\frac{\delta}{\delta J(\bm{r})}\Phi(\bm{r})\frac{\delta}{\delta J^*(\bm{r})}}
\langle\hat{W}[J,J^*]\rangle_t
\nonumber\\
&\quad
=
 e^{-i\frac{\delta}{\delta J}\varPhi\frac{\delta}{\delta J^\dag}}
 e^{i(J^\dag\alpha_t+\alpha_t^\dag J)} e^{-J^\dag\mathcal{F}_tJ},
\intertext{where the formula is written in a matrix representation $\alpha_t^\dag J=\int d^3\bm{r}\,\alpha_t^*(\bm{r})J(\bm{r})$, $J^\dag\mathcal{F}_tJ=\int d^3\bm{r}\,d^3\bm{r}'\,J^*(\bm{r})\times \hat{\mathcal{F}}_t(\bm{r},\bm{r}')J(\bm{r}')$, etc., and $\varPhi(\bm{r},\bm{r}')=\Phi(\bm{r})\delta^3(\bm{r}-\bm{r}')$ is a diagonal matrix. By expressing the Gaussian factor $e^{-J^\dag\mathcal{F}_tJ}$ in its (path-integral) Fourier representation, it becomes easy to apply the functional derivative $e^{-i\frac{\delta}{\delta J}\varPhi\frac{\delta}{\delta J^\dag}}$ to it, and we proceed as}
&\quad
= e^{-i\frac{\delta}{\delta J}\varPhi\frac{\delta}{\delta J^\dag}}
 e^{i(J^\dag\alpha_t+\alpha_t^\dag J)}
\nonumber\\[-1truemm]
&\qquad\ %
{}\times\frac{1}{\Det\mathcal{F}_t}
\int\mathcal{D}^2\gamma\, e^{-\gamma^\dag\mathcal{F}_t^{-1}\gamma}
 e^{- i(J^\dag\gamma+\gamma^\dag J)}
\nonumber\displaybreak[0]
\\
&\quad
=\frac{1}{\Det\mathcal{F}_t}
\int\mathcal{D}^2\gamma\, e^{-\gamma^\dag\mathcal{F}_t^{-1}\gamma}
 e^{i(\gamma-\alpha_t)^\dag\varPhi(\gamma-\alpha_t)}
\nonumber\\[-2truemm]
&\qquad\qquad\qquad\qquad\qquad\qquad\ \ %
{}\times e^{-iJ^\dag(\gamma-\alpha_t)}
 e^{- i(\gamma-\alpha_t)^\dag J}
\nonumber\displaybreak[0]\\
&\quad=
\frac{1}{\Det(1- i\varPhi\mathcal{F}_t)}
 e^{i\alpha_t^\dag(\varPhi^{-1}- i\mathcal{F}_t)^{-1}\alpha_t}
\nonumber\\
&\qquad\ %
{}\times e^{-J^\dag(\mathcal{F}_t^{-1}- i\varPhi)^{-1}J}
 e^{ iJ^\dag(1- i\mathcal{F}_t\varPhi)^{-1}\alpha_t}
 e^{i\alpha_t^\dag(1- i\varPhi\mathcal{F}_t)^{-1}J}.
\end{align}
By putting $J,J^*=0$, we get (\ref{eqn:GaussG}).

\section{Canonical Ensemble}
\label{app:Canonical}
The characteristic functional for the canonical ensemble $\langle\hat{W}[J,J^*]\rangle_N$ defined in (\ref{eqn:CharFuncC}) is to be estimated on the sector with a definite number $N$ of atoms specified by the projection operator $\hat{P}_N$.
It is not easy to carry out such a calculation in a straightforward way, but still, it is possible to obtain the formula for $\langle\hat{W}[J,J^*]\rangle_N$, as demonstrated in \cite{ref:BECCanonical} for the ideal Bose gas in free space.
In this appendix, we derive the formula (\ref{eqn:CharFuncCBessel}) for the canonical ensemble of the ideal Bose gas trapped in a single harmonic potential, in the regime (\ref{eqn:Regime}) relevant in the ordinary experiments.

Observe first that, by noting that 
\begin{equation}
\sum_N\hat{P}_N=1,
\end{equation}
the average in the canonical ensemble $\langle\hat{W}[J,J^*]\rangle_N$ in (\ref{eqn:CharFuncC}) is, in general,  related to that in the grand canonical ensemble $\langle\hat{W}[J,J^*]\rangle_G$ in (\ref{eqn:CharFuncGC}) as 
\begin{align}
\langle\hat{W}[J,J^*]\rangle_G
&=\sum_N\frac{\Tr\{\hat{W}[J,J^*]\hat{P}_Ne^{-\beta(\hat{H}-\mu\hat{N})}\}}{\Tr\{e^{-\beta(\hat{H}-\mu\hat{N})}\}}\nonumber\displaybreak[0]\\
&=\sum_Ne^{\beta\mu N}\frac{\Tr\{\hat{W}[J,J^*]\hat{P}_Ne^{-\beta\hat{H}}\}}{\Tr\{e^{-\beta(\hat{H}-\mu\hat{N})}\}}\nonumber\displaybreak[0]\\
&=\sum_Ne^{\beta\mu N}
\frac{\Tr\{\hat{P}_Ne^{-\beta\hat{H}}\}}{\Tr\{e^{-\beta(\hat{H}-\mu\hat{N})}\}}
\langle\hat{W}[J,J^*]\rangle_N
\nonumber\displaybreak[0]\\
&=\sum_N\langle\hat{P}_N\rangle_G\langle\hat{W}[J,J^*]\rangle_N.
\end{align}

Note that the chemical potential $\mu$ in the grand canonical ensemble $\langle\cdots\rangle_G$ is fixed by imposing 
\begin{equation}
\langle\hat{N}\rangle_G=\sum_{\bm{n}}f(\varepsilon_{\bm{n}})=\bar{N}
\label{eqn:Nbar}
\end{equation}
for a given $\bar{N}$, where $f(\varepsilon)$ is the Bose distribution function defined in (\ref{eqn:Bose}).
For the ideal Bose gas trapped in an isotropic 3D harmonic potential characterized by a trapping frequency $\omega$ (see Sec.\ \ref{sec:SingleGas}), the condition (\ref{eqn:Nbar}) reads \cite{ref:BEC-Stringari}
\begin{equation}
\int_{u_x,u_y,u_z\ge0}d^3\bm{u}\,\frac{1}{e^{-\beta\mu}e^{u_x+u_y+u_z}-1}
=\bar{\sigma}
\label{eqn:MuEq}
\end{equation}
in the limit 
\begin{equation}
\beta\hbar\omega\to0\quad\text{keeping}\quad\bar{\sigma}=(\beta\hbar\omega)^3\bar{N}\quad\text{finite}.
\label{eqn:Scaling}
\end{equation}
There exists a solution $\mu\,(\le0)$ to this equation only when 
\begin{equation}
\bar{\sigma}\le
\tilde{\sigma}=\int_{u_x,u_y,u_z\ge0}d^3\bm{u}\,\frac{1}{e^{u_x+u_y+u_z}-1}=\zeta(3),
\end{equation}
namely, when
\begin{equation}
T\ge T_c
\end{equation}
with $T_c$ defined in (\ref{eqn:Tc}) (but with $\bar{N}$ instead of $N$).
In the other temperature region $T<T_c$, the chemical potential is vanishing $\mu=0$ and the excess number
\begin{equation}
\sigma_0=\bar{\sigma}-\tilde{\sigma}
=\lambda\bar{\sigma}\quad
(T<T_c)
\end{equation}
is attributed to the condensed atoms, with the condensation fraction $\lambda$ presented in (\ref{eqn:Fraction}).

Now, the characteristic functional for the grand canonical ensemble is estimated to be 
\begin{align}
&\langle\hat{W}[J,J^*]\rangle_G
\nonumber\\
&\ %
=e^{-\sum_{\bm{n}}f(\varepsilon_{\bm{n}})|J_{\bm{n}}|^2},\qquad
J_{\bm{n}}=\int d^3\bm{r}\,\varphi_{\bm{n}}^*(\bm{r})J(\bm{r})
\nonumber\displaybreak[0]\\
&\ %
\to\begin{cases}
\displaystyle
\exp\!\left(
-\sigma_0|\mathcal{J}_0|^2-\int_{u_x,u_y,u_z\ge0}d^3\bm{u}\,\frac{|\mathcal{J}_{\bm{u}}|^2}{e^{u_x+u_y+u_z}-1}
\right)\\
\hfill(T<T_c)\\
\displaystyle
\exp\!\left(
-\int_{u_x,u_y,u_z\ge0}d^3\bm{u}\,\frac{|\mathcal{J}_{\bm{u}}|^2}{e^{-\beta\mu}e^{u_x+u_y+u_z}-1}
\right)
\\
\hfill(T\ge T_c)
\end{cases}
\nonumber\displaybreak[0]\\[2.5truemm]
&\ %
\equiv\mathcal{W}_G[\mathcal{J},\mathcal{J}^*]
\end{align}
in the limit (\ref{eqn:Scaling}), where $J$ is scaled as 
\begin{equation}
J_{\bm{n}}=(\beta\hbar\omega)^{3/2}\mathcal{J}_{\bm{u}},\quad
\bm{u}=\beta\hbar\omega\bm{n}.
\label{eqn:ScalingJ}
\end{equation}
As for the factor $\langle\hat{P}_N\rangle_G$, by noting that 
\begin{equation}
\langle e^{-i\theta\hat{N}}\rangle_G
=\prod_{\bm{n}}[1+f(\varepsilon_{\bm{n}})(1-e^{-i\theta})]^{-1},
\end{equation}
we have
\begin{align}
&(\beta\hbar\omega)^{-3}\langle\hat{P}_N\rangle_G
\nonumber\\
&\qquad
=(\beta\hbar\omega)^{-3}\int_{-\pi}^\pi\frac{d\theta}{2\pi}\,\langle e^{i\theta(N-\hat{N})}\rangle_G
\nonumber\displaybreak[0]\\
&\qquad
\to\begin{cases}
\displaystyle
\int_{-\infty}^\infty\frac{d\xi}{2\pi}\frac{e^{i\xi(\sigma-\tilde{\sigma})}}{1+i\xi\sigma_0}=\theta(\sigma-\tilde{\sigma})\frac{1}{\sigma_0}e^{-(\sigma-\tilde{\sigma})/\sigma_0}
\\
\smallskip
\hspace*{55truemm}(T<T_c)
\\
\displaystyle
\int_{-\infty}^\infty\frac{d\xi}{2\pi}\,e^{i\xi(\sigma-\bar{\sigma})}=\delta(\sigma-\bar{\sigma})\hfill(T\ge T_c)
\end{cases}
\nonumber\\[1.5truemm]
&\qquad
\equiv\mathcal{K}(\sigma),
\end{align}
with $\sigma=(\beta\hbar\omega)^3N$.
The characteristic functional for the canonical ensemble in the continuum limit
\begin{equation}
\langle\hat{W}[J,J^*]\rangle_N\to\mathcal{W}_\sigma[\mathcal{J},\mathcal{J}^*]
\end{equation}
is then available by inverting the relation 
\begin{equation}
\mathcal{W}_G[\mathcal{J},\mathcal{J}^*]
=\int_0^\infty d\sigma\,\mathcal{K}(\sigma)\mathcal{W}_\sigma[\mathcal{J},\mathcal{J}^*].
\label{eqn:GandC}
\end{equation}

Above the critical temperature $T\ge T_c$, it is just given by
\begin{align}
\mathcal{W}_\sigma[\mathcal{J},\mathcal{J}^*]
&=\mathcal{W}_G[\mathcal{J},\mathcal{J}^*]\biggr|_{\bar{\sigma}=\sigma}
\nonumber\\
&=\exp\!\left(
-\int_{u_x,u_y,u_z\ge0}\!\!\!\!\!\!d^3\bm{u}\,\frac{|\mathcal{J}_{\bm{u}}|^2}{e^{-\beta\mu}e^{u_x+u_y+u_z}-1}
\right)
\nonumber\\
&\hspace*{54truemm}
(T\ge T_c),
\label{eqn:CharCcontAbove}
\end{align}
where $\mu$ is fixed by (\ref{eqn:MuEq}) with $\sigma$ in place of $\bar{\sigma}$.
Below the critical temperature $T<T_c$, on the other hand, the relation (\ref{eqn:GandC}) is essentially the Laplace transformation,
\begin{align}
&\int_0^\infty d\sigma'\,e^{-\sigma'/\sigma_0}\mathcal{W}_{\sigma'+\tilde{\sigma}}[\mathcal{J},\mathcal{J}^*]
\nonumber\\
&\ \ %
=\sigma_0e^{-\sigma_0|\mathcal{J}_0|^2}
\exp\!\left(
-\int_{u_x,u_y,u_z\ge0}d^3\bm{u}\,\frac{|\mathcal{J}_{\bm{u}}|^2}{e^{u_x+u_y+u_z}-1}
\right)
\nonumber\\
&\hspace*{71truemm}(T<T_c),
\end{align}
which is inverted, by noting a formula for the Bessel function
\begin{equation}
J_\nu(z)=\left(\frac{z}{2}\right)^\nu\int_{C_B}\frac{ds}{2\pi i}\,s^{-\nu-1}e^{s-z^2/4s},
\end{equation}
to yield
\begin{align}
\mathcal{W}_\sigma[\mathcal{J},\mathcal{J}^*]
={}&J_0(2\sqrt{\sigma-\tilde{\sigma}}|\mathcal{J}_0|)
\nonumber\\
&{}\times\exp\!\left(
-\int_{u_x,u_y,u_z\ge0}d^3\bm{u}\,\frac{|\mathcal{J}_{\bm{u}}|^2}{e^{u_x+u_y+u_z}-1}
\right)
\nonumber\\
&\hspace*{50truemm}(T<T_c).
\label{eqn:CharCcontBelow}
\end{align}
Equations (\ref{eqn:CharCcontAbove}) and (\ref{eqn:CharCcontBelow}) are presented in (\ref{eqn:CharFuncCBessel}), keeping in mind the limit (\ref{eqn:Scaling}) and the scaling (\ref{eqn:ScalingJ}).

\section{Condensation Fraction and Purity}
\label{app:Purity}
Let us estimate $\Tr\{\hat{\mathcal{F}}'^2\}$ and $\Tr\{\hat{\mathcal{F}}'^4\}$, which control the average spectrum $S_t(\bm{k})$ and the covariance $C_t(\bm{k},\bm{k}')$ in Sec.\ \ref{sec:IntCond}\@.
We start by looking at $\Tr\{\hat{\mathcal{F}}'\}$.
For the ideal Bose gas trapped in a single harmonic potential, setup in Sec.\ \ref{sec:SingleGas}, we have \cite{ref:Ketterle-BECHarmonic-PRA}
\begin{align}
N&=\sum_{\bm{n}}\frac{1}{e^{\beta(\varepsilon_{\bm{n}}-\mu)}-1}
\nonumber\\
&=\sum_{\bm{n}}
\sum_{j=1}^\infty e^{-j\beta(\varepsilon_{\bm{n}}-\mu)}
\nonumber\\
&=\sum_{j=1}^\infty z^j
\frac{1}{(1-e^{-j\beta\hbar\omega})^3},
\intertext{where $z=e^{\beta\mu}$.
By splitting the occupation of the ground state,}
&=\frac{z}{1-z}
+\sum_{j=1}^\infty z^j
\left(
\frac{1}{(1-e^{-j\beta\hbar\omega})^3}-1
\right).
\intertext{Now, since $\beta\hbar\omega\ll1$ in the regime (\ref{eqn:Regime}), we approximate it by \cite{ref:Ketterle-BECHarmonic-PRA}}
&\simeq\frac{z}{1-z}
+\frac{1}{(\beta\hbar\omega)^3}\sum_{j=1}^\infty\frac{z^j}{j^3}
\nonumber\displaybreak[0]
\\
&=\frac{z}{1-z}
+\frac{1}{(\beta\hbar\omega)^3}
g_3(z),
\end{align}
where
\begin{equation}
g_n(z)=\sum_{j=1}^\infty\frac{z^j}{j^n}.
\end{equation}
Therefore, the condensation fraction is given by
\begin{equation}
\lambda
=\frac{1}{N}\frac{z}{1-z}
\simeq1-\frac{g_3(z)}{(\beta\hbar\omega)^3N}
=1-\left(\frac{T}{T_c}\right)^3\frac{g_3(z)}{\zeta(3)},
\label{eqn:lambda}
\end{equation}
where $T_c$ is defined in (\ref{eqn:Tc}).
This is nothing but (\ref{eqn:Fraction}).
Remember the condition for the chemical potential (\ref{eqn:CondMu}) and $g_3(1)=\zeta(3)$.

Quite similarly, the purity of the gas is estimated as
\begin{align}
&\sum_{\bm{n}}\frac{1}{(e^{\beta(\varepsilon_{\bm{n}}-\mu)}-1)^2}
\nonumber\\
&\qquad
=\sum_{j=1}^\infty (j-1)z^j
\frac{1}{(1-e^{-j\beta\hbar\omega})^3}
\nonumber\displaybreak[0]
\\
&\qquad
=\left(\frac{z}{1-z}\right)^2
+\sum_{j=1}^\infty(j-1)z^j
\left(
\frac{1}{(1-e^{-j\beta\hbar\omega})^3}-1
\right)
\nonumber\displaybreak[0]
\\
&\qquad
\simeq\left(\frac{z}{1-z}\right)^2
+\frac{1}{(\beta\hbar\omega)^3}
[g_2(z)-g_3(z)],
\end{align}
\begin{widetext}
\noindent
i.e.,
\begin{equation}
\Tr\{\hat{\mathcal{F}}^2\}
\simeq\lambda^2+\frac{1}{N}\left(\frac{T}{T_c}\right)^3\frac{g_2(z)-g_3(z)}{\zeta(3)}.
\label{eqn:F2}
\end{equation}
Note that $g_2(1)=\zeta(2)=\pi^2/6$.
Furthermore,
\begin{align}
\sum_{\bm{n}}\frac{1}{(e^{\beta(\varepsilon_{\bm{n}}-\mu)}-1)^4}
&=\sum_{j=1}^\infty (j-1)(j-2)(j-3)z^j
\frac{1}{(1-e^{-j\beta\hbar\omega})^3}
\nonumber\displaybreak[0]\\
&\simeq\left(\frac{z}{1-z}\right)^4
+\frac{1}{6(\beta\hbar\omega)^3}
\left(
\frac{z}{1-z}+6\ln(1-z)
+11g_2(z)-6g_3(z)]
\right),
\end{align}
and hence,
\begin{equation}
\Tr\{\hat{\mathcal{F}}^4\}
\simeq\lambda^4
+\frac{1}{6\zeta(3)N^2}
\left(\frac{T}{T_c}\right)^3
\left[
\lambda\left(
1+6\frac{1-z}{z}\ln(1-z)
\right)
+\frac{1}{N}[11g_2(z)-6g_3(z)]
\right].
\label{eqn:F4}
\end{equation}
\end{widetext}


\begin{thebibliography}{10}
\bibitem{note:Purity}
The variance of the total number of atoms, $S_t(0)$, should be strictly zero in
  the case of canonical ensemble, since the total number of atoms is precisely
  fixed, while $S_t(0)$ in (\ref{eqn:Sasymp}) gives $O(1/N)$. It is consistent
  within the validity of the approximation employed to compute the
  characteristic functional $Z_t^N[\Phi]$.
\end{thebibliography}

\end{document}